\documentclass[10pt, conference, compsocconf]{IEEEtran}

\usepackage[utf8]{inputenc}
\usepackage[T1]{fontenc}
\usepackage{microtype}
\usepackage{comment}

\usepackage{amsmath}
\usepackage{amsthm}
\usepackage{amssymb}
\usepackage{xcolor}
\usepackage{graphicx}
\usepackage{url}
\usepackage{algorithmicx}
\algblock{Input}{EndInput}
\algnotext{EndInput}
\algblock{Output}{EndOutput}
\algnotext{EndOutput}
\usepackage{algorithm}
\usepackage[noend]{algpseudocode}
\usepackage{booktabs}
\usepackage{tabularx}
\usepackage{colortbl}
\usepackage{xspace}
\usepackage{soul}
\usepackage{algorithm,algpseudocode}
\usepackage{multirow}
\usepackage{tabularray}
\usepackage[normalem]{ulem}
\usepackage{multirow}
\usepackage[nocompress]{cite}
\usepackage{makecell}
\usepackage{multirow}
\usepackage{array}

\usepackage{diagbox}

\theoremstyle{definition}
\newtheorem{definition}{Definition}

\newcommand{\explanation}{\emph{explanation}\xspace}
\newcommand{\explanandum}{\emph{explanandum}\xspace}
\newcommand{\explainee}{\emph{explainee}\xspace}
\newcommand{\exen}{\emph{SmartEx}\xspace}
\newcommand{\smartEnv}{smart environments\xspace}

\newcommand{\ConExpComp}{\textit{Context-Aware Explanation Generator}\xspace}
\newcommand{\AlExpComp}{\textit{Algorithmic Explanation Generator}\xspace}
\newcommand{\ConMangComp}{\textit{Context Manager}\xspace}

\newcommand{\Find}{\textbf{find }}
\newlength\myindent
\newcommand{\funcall}[1]{\text{#1}}

\renewcommand{\baselinestretch}{0.95}

\begin{document}

\title{SmartEx: A Framework for Generating User-Centric Explanations in Smart Environments}

\author{ 
\IEEEauthorblockN{Mersedeh Sadeghi, Lars Herbold, Max Unterbusch, Andreas Vogelsang \\ }
 \IEEEauthorblockA{University of Cologne, Cologne, Germany\\
   sadeghi@cs.uni-koeln.de, lars.herbold@outlook.de, munterbu@smail.uni-koeln.de, vogelsang@cs.uni-koeln.de }
}

\maketitle

\begin{abstract}
Explainability is crucial for complex systems like pervasive smart environments, as they collect and analyze data from various sensors, follow multiple rules, and control different devices resulting in behavior that is not trivial and, thus, should be explained to the users. The current approaches, however, offer flat, static, and algorithm-focused explanations. User-centric explanations, on the other hand, consider the recipient and context, providing personalized and context-aware explanations. To address this gap, we propose an approach to incorporate user-centric explanations into smart environments. We introduce a conceptual model and a reference architecture for characterizing and generating such explanations. Our work is the first technical solution for generating context-aware and granular explanations in smart environments. Our architecture implementation demonstrates the feasibility of our approach through various scenarios. 
\end{abstract}

\IEEEpeerreviewmaketitle

\section{Introduction}\label{Sec:Intro}
Explainability has become a vital research topic due to the recent progress in machine learning, autonomous systems, complex decision-making, and smart cyber-physical systems~\cite{adadi2018peeking,arrieta2020explainable,nunes2017systematic,tjoa2020survey,jakobi2018evolving,minh2022explainable,puiutta2020explainable}. Users can lose trust in a system if they feel they have no control and if the system's behavior does not meet their expectations, leading to misuse or rejection~\cite{lim2009and, meerbeek2014impact}. 
Many studies indicate that providing explanations is a viable solution because it enables users to develop a more accurate mental model of the system, allowing them to anticipate and interpret its behavior more effectively~\cite{winikoff2017towards,gunning2017explainable, sauer2007automation, jakobi2018evolving}.

The advancements of smart devices, cloud computing, mobile networks, and pervasive computing have resulted in highly complex ecosystems~\cite{vermasan2014internet}. These ecosystems gather extensive data on the environment and users' actions, goals, and preferences, facilitating context-aware decision-making and autonomous activities. The ultimate goal of this pervasive intelligence is to enhance user satisfaction and comfort. Nevertheless, the autonomous and automated nature of these systems can often lead to confusion and frustration among users~\cite{jakobi2018evolving}. Subsequently, there is a need to provide explanations for users at runtime to clarify the cause, purpose, and benefit of the system's actions.

We've identified issues with current explanation approaches in the \smartEnv domain. Firstly, many of them focus solely on algorithmic explanations~\cite{abdul2018trends,liao2021human, mueller2021principles}, which describe system behavior in terms of algorithms and processes~\cite{waskan2011mechanistic,doshi2017towards,glennan2002rethinking, liao2021human}. These explanations often unravel the planning or a causal chain of events, that are challenging for end-users to comprehend~\cite{miller2019explanation,lipton1990contrastive, ehsan2020human}. Furthermore, algorithmic explanations neglect the user's context. They provide uniform explanations, overlooking other explanation desiderata besides describing the inner workings of a system, such as persuasion, learning, or assignment of blame~\cite{lombrozo2006structure,wilkenfeld2015inference}.
The algorithmic explanation constitutes an objective explanation, that only considers the problem to be explained. However, the other desiderata require a more subjective explanation, that considers the context in addition to the problem. Hence, the user-centric explanation should encompass contextual elements, reasoning, and information beyond cause-effect relations.~\cite{miller2019explanation, mueller2021principles}. Finally, existing approaches for explanation generation in \smartEnv often suffer from a lack of personalization. Bunt~et~al.'s empirical study~\cite{bunt2012explanations} shows differing perceptions of explanation attributes like completeness, soundness, and complexity. This suggests that a one-fits-all explanation is insufficient, and there is a need for personalized and granular explanations that consider individual differences. 
Despite wide agreement on the importance and essential attributes of user-centric explanations, we lack approaches for developing smart environment systems that can generate them. 

To address the aforementioned shortcomings, we propose building an explanation layer on top of a smart environment system that constructs user-centric explanations by leveraging the capabilities inherent in such systems (e.g., physical sensors and devices, data collection and processing components, and knowledge extraction mechanisms). Our approach offers the following contributions:

\begin{itemize}
\item We define a model for user-centric explanations in \smartEnv.

\item We present a reference architecture for a user-centric explanation generation engine, identifying event causes within a rule-based system and enhancing explanations with pertinent context to deliver personalized, context-aware explanations in natural language.

\item To assess the feasibility of our solution, we have implemented the proposed reference architecture.
\end{itemize}

\section{Related Work}\label{Sec:Related}

Ambient smart environments proactively support people in their daily lives~\cite{cook2009ambient}.  According to Huang~et~al.~\cite{huang2015supporting}, a mismatch between a user's mental model and a system's logical interpretation can result in program failure and reduced perceived ease of use. Using explanations effectively addresses the issue by enhancing system transparency, improving user mental models, and ultimately boosting trust and usability~\cite{winikoff2017towards,sauer2007automation,jakobi2018evolving}. Many studies emphasize the significance of experiential, explorative, contextual, and timely explanations to improve user interpretability; adding a human touch can increase social presence~\cite{lakkaraju2017interpretable, mueller2021principles, brezillon1997joint, li2015hedonic}. Miller~\cite{miller2019explanation}, Hoffman~et~al.~\cite{hoffman2017explaining, hoffman2017explainingp2}, Klein~\cite{klein2018explaining}, and Mittelstadt~et~al.~\cite{mittelstadt2019explaining} in their studies have overviewed decades of works on social sciences for explanation attribution. According to their finding, effective explanations must be user-centric, i.e., cater to users' epistemic states, offering both ``everyday'' and scientific explanations. Paes~\cite{paez2019pragmatic} discusses that explanations need to convey both a pragmatic and naturalistic account of understanding, and Gregor and Benbasat~\cite{gregor1999explanations} reason that explanations for end-users must be context-based. 

Moving to Software Engineering for Explainable Smart Environments, we can refer to some related works. FORTNIoT~\cite{zhao2021understanding} is a smart home framework that offers a self-sustaining prediction model and simulations for forecasting rule occurrences in smart homes. Additionally, it provides explanations for future entity changes, helping users identify triggering rules and their reasons. Unlike our research, FORTNIoT mainly focuses on forecasting events and provides limited explanations that show cause-effect relations among rules, lacking contextual and personalization aspects.
Houz{'e}~et~al.~\cite{houze2022generic} developed an explainable smart home framework, comprising explanatory components for specific types of sensors and smart devices. Like our approach, they see smart home dynamics as predicates and trace events to pinpoint which predicates support the questioned proposition, constituting the explanation. However, their architecture solely deals with simple rule-based systems where individual objects have their rules (e.g., a thermostat turning off if the temperature exceeds a threshold) and don't delve into complex rules involving group devices, predicates, and events. Moreover, objects must adhere to standard interfaces to integrate into their framework, which limits their effectiveness due to the wide diversity of smart devices.
Authors in a set of works~\cite{kordts2021towards, burmeister2017smart} focused on developing a loosely coupled mechanism for generating explanations. They introduced a description language for smart devices and a mediator acting as middleware to handle the diversity of sensors and smart objects. Their unique contribution lies in incorporating information about device functions into the description, enabling smart devices to provide self-explanations. In contrast to our approach, their work primarily aims at offering tutorials to help users better operate devices. In our work, our primary focus is on enabling intelligent systems to self-explain automated actions and decisions to enhance human-computer interactions and engagement for end-users.
In a related vein, there is a substantial body of research pursuing a similar objective in a broader application domain, like cyber-physical systems. Blumreiter et al.~\cite{blumreiter2019towards} introduced MAB-EX (Monitor, Analyze, Build, Explain), a reference architecture for generating explanations based on the MAPE-K framework for self-adaptive systems\cite{Kephart2003}. In their system, an explanation takes the form of a behavioral model of the system, capturing causal relationships between events and system responses. The leaves in their cause tree are linked to static explanations that can ultimately be presented to the user. While the authors discussed the potential for automated explanation creation, their initial demonstration relied on manual explanation generation.

To conclude this section, we can observe that the focus in explainability domains has mainly revolved around algorithmic explanations. These explanations mainly hinge on causal relations, presented as diagrams, charts, or stats, which can be challenging for end-users. Hence, there is a growing imperative to shift towards user-centric~\cite{lipton2018mythos, gilpin2018explaining}. In our work, we start with causal explanation generation but then extend it with contextual elements by utilizing the inherent capabilities of a smart environment. Indeed, smart environments, comprising interconnected sensors, devices, and mechanisms, offer valuable resources for user-centric explanations. Despite the potential of context-awareness in enhancing various aspects of ambient environment systems, such as personalization, location-based services, and activity recognition~\cite{neuhofer2015smart,alletto2015indoor,baresi2018tde,ehatisham2020opportunistic}, explainability in smart environments is often overlooked, as highlighted in related works~\cite{houze2022generic,kordts2021towards}. Our research aims to bridge this gap by offering a practical framework for creating user-centric explainable systems.

\section{User-Centric Explanations Conceptualization}\label{Sec:expModel}

\subsection{Motivating Scenario}\label{subsec:Motiv}
We devised a scenario inspired by a collection of situations that need to be explained in an interactive smart environment proposed in~\cite{sadeghi2021cases}. The scenario involves three users---Alice, Bob, and Dana---interacting in an office building equipped with smart devices and Smart Office Manager (SOM) software, controlling various intelligent functionalities. Bob has set up a rule to mute the TV automatically whenever a meeting occurs in rooms located near the kitchen. Sometime later, Alice tries to play music during lunch, but the TV mutes itself, leading to her frustration. Despite her attempts to unmute it, the TV keeps muting itself. Alice needs an explanation for this behavior. Bob sees Alice struggling with the TV but does not realize he set a rule to mute it during nearby meetings, especially since he is unaware of the meeting in Room 1. Now, he needs an explanation too. On another occasion, Dana, a visitor, is waiting in the kitchen for her host Chuck, and experiences the TV suddenly going silent. Despite not actively watching the TV, the abrupt change in the TV audio left Dana confused, particularly because she is alone in the kitchen and has not interacted with any system. As a result, she feels the need for an explanation.

In this scenario, three individuals find themselves in a perplexing situation. However, as we will elaborate in the upcoming section, each individual requires a personalized explanation distinct from the others. Moreover, the explanation must be easily accessible to them through the same smart infrastructure they use for managing the intelligent environment. This pervasive provision of explanation is crucial for facilitating user interaction.

\subsection{Explanation Conception}\label{subsec:formal}

\begin{table*}
\centering
\footnotesize
\caption{Example of Algorithmic and Semantic Constructs in our model}
\label{table:events}
\begin{tabularx}{\textwidth}{@{}p{5.3cm}X@{}} 
\toprule
\textbf{Explanation Construct }& \textbf{Example} \\
 \midrule
Algorithmic Explanation Constructs \newline (for causal explanation)    & 

($\chi_r$: $Rule_k$ is fired) $\rightarrow$ ($\chi_s$: TV is Mute)  \newline
($\chi_p$~: Meeting M1 is going on in Room 1) $\wedge$ ($\chi_q$: TV is On) $\rightarrow$  ($\chi_r$: $Rule_k$ is fired)
\\ \hline

Contextual Explanation Constructs \newline (for context-based explanation) &

($\chi_i$ : Bob has set rule $Rule_k) \wedge$ \newline ($\chi_j$ : $Rule_k$ states that if TV is on while a meeting is taking place in rooms 1, 2, or 3, then mute TV)\\
\bottomrule
\end{tabularx}
\end{table*}

Before presenting our explanation formalization in Definition~\ref{def:psi}, we provide a brief overview of the notations used throughout this work. The term \textit{Smart Environment System S} defines modern smart homes, offices, and buildings, as having interconnected sensors and intelligent devices communicating with each other and users. They typically possess autonomous perception, reasoning, and action capabilities in their environment. Additionally, the term \textbf{Explanandum}~$\boldsymbol{\phi}$ refers to the state or behavior of a system~\emph{S} that an \textbf{Explainee} requests an explanation for. An \textbf{\textit{Explainee}}, in our study, is a human who interacts with the system~\emph{S}.
Building on the causal explanation framework introduced in~\cite{halpern2005causes}, we broaden its application to encompass user-specific and contextual factors overlooked in the generic approach.
The original formulation proposes a causal explanation of a phenomenon \emph{P} through a conjunction of primitive \emph{Events}. This method offers the advantage of a simple and generic structure for composing explanations by identifying a sequence or combination of events that led to the occurrence of \emph{P}. However, the main drawback is that it lacks consideration for the user and contexts resulting in a mechanistic explanation that may not address the user's needs.

\noindent Therefore, our objective is twofold. Firstly, we aim to apply the concept of the generic causal explanation formalization presented in the existing literature to the context of \smartEnv. Secondly, we intend to extend this concept by integrating contextual aspects. To facilitate this, we introduce the notion of \textbf{Explanation Constructs} denoted as \textbf{\emph{X}}.

In our model, the \emph{Explanation Constructs} encompass a set of specifications, facts, propositions, and events related to both the internal elements \emph{S} and the external world which includes the user's states and external systems, such as a list of smart objects managed by \emph{S}, a general description of such devices, their current and past states, sequences of actions and events, a set of rules and specifications, a set of contextual variables (e.g., user's roles) and their current values.
\emph{Explanation Constructs} hence are among the data that are commonly accessible and processable in modern smart environments through APIs, documentation, and logs of events provided either by \emph{S} itself or external systems. Finally, to proceed with the \emph{Explanation} definition, we can broadly classify \emph{Explanation Constructs} into two categories as shown in Table~\ref{table:events}.\

The first category is \textbf{Algorithmic Explanation Constructs (AEC)}, which consists of facts, propositions, specifications, or events that describe the logic or mechanism behind an action/output of a machine/program. The second category is \textbf{Contextual Explanation Constructs (CEC)}, which includes facts, propositions, or specifications that provide information related to an action/output of a machine /program or an \textit{AEC}. By incorporating both AECs and CECs in an explanation, we can create personalized and context-aware explanations. It goes beyond the traditional causal explanations that only describe the underlying mechanism of an \explanandum. %
We define the notion of \emph{Explanation} in the domain of an \smartEnv in Definition~\ref{def:psi}.

\begin{definition}[\textbf{Explanation} $\boldsymbol\Psi$ ]\label{def:psi} 
For a given $\phi$, there is a $\Psi$ that is a conjunction of some explanation constructs: 
$$ \Psi = (\chi_1 \in X) \wedge  \ldots \wedge  (\chi_k \in X)\hspace{0.5em}  where \hspace{0.5em} X = \{\mathit{AEC} \cup \mathit{CEC}\} $$
\end{definition}

Hence, given the nature of \textit{X} and Definition~\ref{def:psi}, an \textit{explanation} is interpreted as a piece of causal and contextual information that describes the algorithmic and contextual reasons behind an \explanandum. Therefore, it is aligned with the user-centric explanation characteristics remarked earlier. More concretely, an explanation is partly constructed by a causal explanation for the system's behavior captured through \emph{AEC}. As shown in Table~\ref{table:events}, for the motivating scenario given in Section~\ref{subsec:Motiv}, where the \explanandum is : [\textit{$\phi$: why is the TV muted?}], the higher level \emph{AEC} can be framed as: [\textit{the TV is muted as a direct result of firing $Rule_k$}]. If one desires to explore the causation path in more depth\footnote{The \explanandum may also be: [\textit{$\phi$: why has this rule been fired?}]}, then \emph{AEC} can also include: [\textit{because a meeting in a nearby room is going on} (\textit{$\chi_p$} in Table~\ref{table:events}) \textit{and TV is on} \textit{(\textit{$\chi_q$}} in Table~\ref{table:events})] since such proposition implies that $Rule_k$ (\textit{$\chi_r$} in Table~\ref{table:events}) fires: [$ \chi_p\wedge \chi_q \rightarrow \chi_r$]. Furthermore, the complementary part of the explanation in our model is the \emph{CEC} that provides further meta-information. In our example, this includes the rule description itself and a rationale as to why such rules exist (see \textit{$\chi_j$} and \textit{$\chi_i$} in Table~\ref{table:events}). 

Our explanation model includes two further essential elements: \emph{Views} and an \emph{Inference Function}. Together, these elements enhance the personalization and context-awareness of an explanation going beyond a one-fits-all approach. 
By different \emph{Views}, we can tailor an explanation to specific user needs and contexts. The \emph{Inference Function}  determines how the system selects a particular \emph{View} to be shown to a particular user based on relevant contextual information. \emph{Transformation Function} ensures that the explanation is understandably presented to the end-user by transforming the formal model of the explanation to a natural language representation.

\begin{definition}[Views of $\Psi$]\label{def:Rep} Let $\Psi$ be the \explanation for $\phi$. There are \textit{n} different  $\mathit{Views}=\{V_1, V_2,\ldots , V_n$\} of $\Psi$, which are combinations of $\mathit{AEC}, \mathit{CEC} \subseteq \Psi$ representing different partial \emph{explanations} for $\phi$. Note that the most detailed and comprehensive view ($V_{\mathit{full}}$) includes all $\mathit{AEC}, \mathit{CEC} \subseteq \Psi$, which essentially means $V_{\mathit{full}}=\Psi$.
\end{definition}

\begin{definition}[Inference Function]\label{def:inference}
There exist some mechanisms (e.g., mapping, inference, or rule engines) that operate by considering a specific \explanandum and all pertinent contexts to deduce the \textit{most suitable} $V_i \in \mathit{Views}$ for presentation to a particular \explainee within a specific situation.
\end{definition}

\begin{definition}[Transformation Function]\label{def:trans}
A function that translates a view $V_i\in \mathit{Views}$ (and its contained AEC and CEC) into a natural language representation.
\end{definition}

In our toy example, we say that $\Psi= \{ \chi_p \wedge \chi_q \wedge \chi_r \wedge \chi_s \wedge \chi_i \wedge \chi_j\}$ (see Table~\ref{table:events}). Given the epistemic state of Bob, who has set the rule \textit{$R_k$}, a relevant explanation is \textit{\textsf{Inference}~(\explainee, Contexts, $\Psi$)~=~$V_1$}, where $V_1$~=~$\{ \chi_p \wedge \chi_q \}$ (see Table~\ref{table:events}). That is because Bob is aware of the existence of \textit{$Rule_k$}. Thus by $\chi_p$ and $\chi_q$, he can infer that the rule has been triggered. This representation is ultimately presented to Bob as \textit{\textsf{Transform}~($V_1$)~=~\{Because Meeting M1 in Room R1 is going on and TV is turned on\}}. However, to personalize the same $\Psi$ for Alice, as a colleague of Bob, there is a need to additionally explain the rule. Therefore, given \textit{\textsf{Inference}~(\explainee, Contexts, $\Psi$)}, a relevant explanation could be $V_2$~=~$\{ \chi_p \wedge \chi_q \wedge \chi_j \}$. Similarly, for Dana, who is a guest visitor and Bob prefers to avoid revealing the underlying details for her, yet another representation $V_3$=$\{ \chi_s \wedge \chi_i \wedge \chi_r \}$ must be synthesized to customize the original explanation. It is then offered to Dana as \textsf{Transform}($V_3$)~=~\{\textit{TV is muted because Bob has set a rule which has been just fired.}\}. 

\section{A User-centric Explanation Engine}\label{Sec:arch}

\begin{figure}
  \centering
\includegraphics[width=\linewidth]{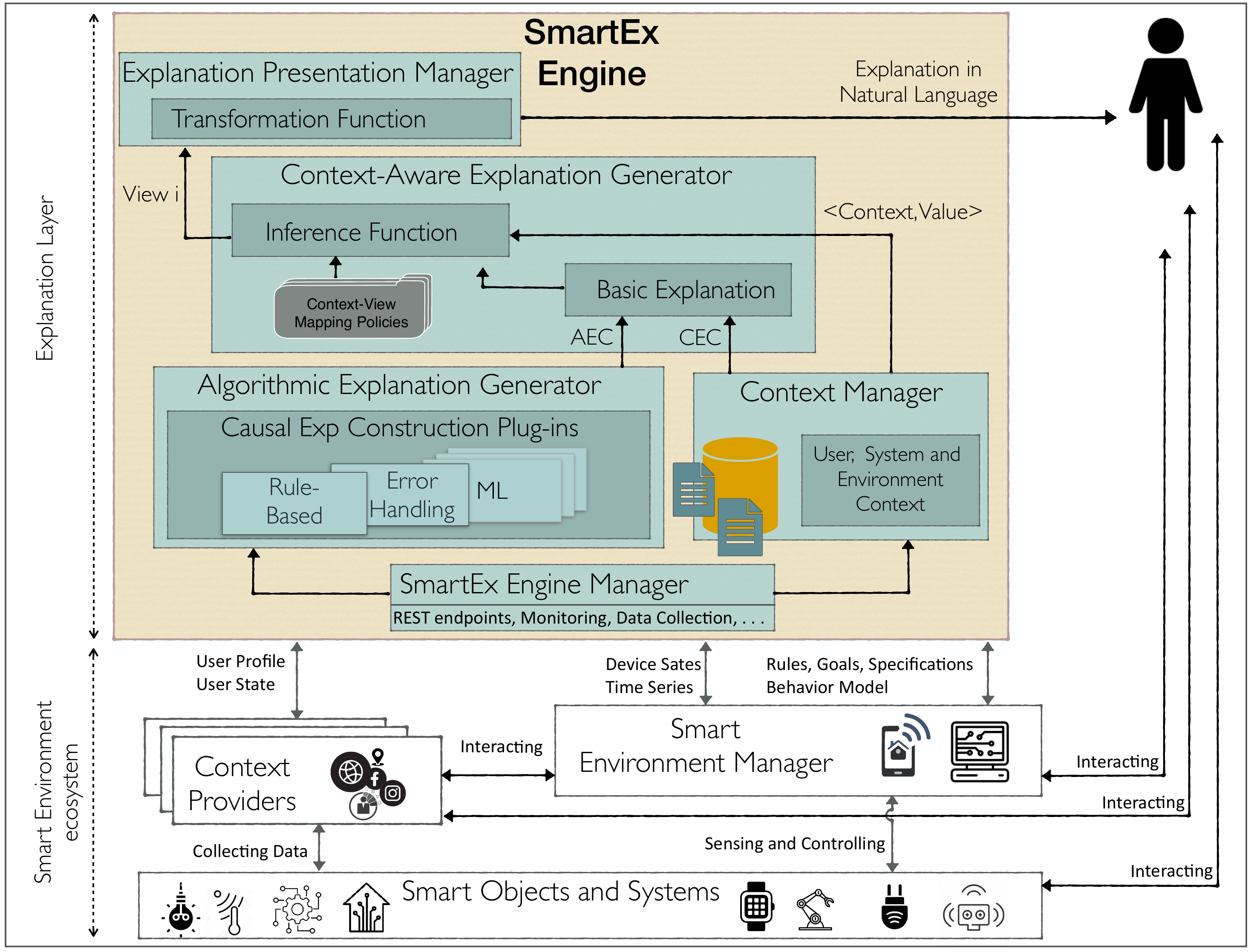}
  \caption{Reference architecture for the Smart Explanation Engine (\exen)}
  \label{fig:arch}
  \vspace{-1em}
\end{figure}

Figure~\ref{fig:arch} shows the reference architecture for the Smart Explanation Engine (\exen), a RESTful service to extend typical smart environment systems to an explainable ecosystem. It decouples the explanation generation process from the core intelligent system, which includes components like the smart environment management system and home automation hub, as well as intelligent device services and applications. 

To create user-centric explanations, the \ConExpComp component needs to have the current usage contexts supplied by the \ConMangComp and the \textit{algorithmic explanation} which is the output of the \AlExpComp component. The \ConMangComp collects and preprocesses contextual elements. Some, like the \textit{Time} context, can be obtained directly on request, while others, such as relatively stable user details (e.g., name, age), are stored locally for future access. Additionally, it can utilize external context providers (e.g., smart environment APIs, social networks) for more complex and dynamic contexts. The supplied contextual elements will then be used by the \textit{Inference Function}. It is responsible for mapping the current values of the relevant contexts to a particular \textit{View} (see Definitions~\ref{def:Rep} and~\ref{def:inference}) among the set of \textit{Views} known by the system. Such a view will ultimately be translated into natural language, which is understandable for a human interacting with the system.

The implementation of the \AlExpComp and the set of \textit{Views}, depends on intelligent methods deployed in the systems, such as various decision models (e.g., rule-based, hidden Markov model, decision tree), recommendation systems (collaborative and content-based filtering), path planning, or ML systems. 
In this work, we are targeting the Rule-based systems (see Section~\ref{subsec:rul}), which are among the most popular approaches for building smart environments and context-aware systems \cite{perera2013context,lim2010toolkit}. However, new plug-ins of explanation generations tailored for other intelligent methods can be integrated into the system in future works.

\subsection{Explanations for Rule-based Systems}\label{subsec:rul}

Each smart object in a typical rule-based system in the \smartEnv has \textit{Actions}, which are operations performed by the object, and \textit{Properties} that describe the internal states of the object. A business \textit{Rule} in its most generic form can be defined as an expression: \textbf{If} \textit{Precondition}, \textbf{then} \textit{Action}. %
In smart environment systems, rules serve as proxies to trigger specific \textit{Actions} for smart objects when certain preconditions (e.g., specific property values or contextual factors) are fulfilled.

We can conceptualize a rule-based smart environment system~$S$ as a knowledge graph, where each node represents a conceptual or physical element, such as a smart object, its actions and properties, rules, and environmental contexts within the system~$S$. Starting from an \textit{Action} node in this graph, multiple paths unfold, representing plausible causes for the execution of that specific action. For instance, in our motivating scenario, the ``mute TV'' action may be issued due to a fired rule (from a set of rules $R_1,\ldots, R_n$), a direct command via an API, or due to remote control signals. Each of these possibilities forms a path in the graph, with the root node being ``mute TV''. Each rule ($R_1,\ldots, R_n$) involves distinct preconditions, combining various smart object properties, user actions, and environmental context. In reality, among all these paths, only one path exists that represents the real cause of an action. \footnote{We assume no duplicate rules and no rule conflicts.} This unique path represents the cause-and-effect chain. In our example, this path comprises preconditions (TV is on and there is a meeting nearby), leading to the fired rule (say, $R_1$), resulting in the ``mute TV'' action.

Our framework has information about the rules of a system including their preconditions and actions. By comparing the rule descriptions with observed events extracted from system logs, we can construct a cause-and-effect chain for an action. To achieve this, our system utilizes a mechanism (see Section~\ref{subsec:algorithmic}) to determine the causal path behind a system behavior. After constructing the causal explanation component, \exen further refines the explanation considering relevant contextual factors (see Section~\ref{subsec:context}).

\subsubsection{\textbf{Algorithmic Explanation}}
\label{subsec:algorithmic} 
Algorithm~\ref{algo:path} shows how we identify the cause-and-effect path discussed above. It takes two inputs. The first is the \textit{Action} to be explained (i.e., the \explanandum). The second is a set of \emph{Explanation Constructs} $X$, which include a list of smart objects managed by $S$, a general description of such devices, their current and past states, sequences of actions, and events, i.e., the log of $S$ from $m$ minutes ago until the time the explanation was requested. The output of the algorithm is either the unique path that results in the \explanandum or \textit{Null} if no such path exists.

In essence, the path corresponds to the rule that has been triggered, resulting in the execution of the \explanandum (\textit{FiredRule} in Alg.~\ref{algo:path}), along with all preconditions and actions associated with that rule. To ascertain the output, the algorithm initiates by extracting and removing $R$ from the set $X$. $R$ represents the collection of rules established by users. The remaining constructs in $X$ are then sorted in order from $T_{\mathit{now}}$ (when the request for generating an explanation is made) back until \textit{$T_m$}. The variable $m$ is configurable and determines the extent to which the algorithm examines past events.%

\begin{algorithm}
\scriptsize
    \caption{Find the cause path}
    {
    \begin{algorithmic}[1]
    \Procedure{\funcall{Find the cause path $P$}}{}\\ 
    \textbf{input:} {$X$: Set of all explanation constructs observed by the system (i.e., rules, actions, preconditions, logs of events, etc.), $explanandum$: the explanandum} \\  
    \textbf{output} {path: $P$}

    \State $P \gets \emptyset$
     \State $\mathit{FiredRule} \gets \emptyset$
     \State $R \gets$ Set of rules extracted from $X$ 
     \State $X \gets X-R$
  
    \State {\textbf{sort}  ${x} \in X$ in reverse chronological order from ${t_{\mathit{now}}}$ to $t_m$} 
        \State $\mathit{CandidateRules} \gets r  \hspace{0.5 mm} \mathit{where} \hspace{0.5 mm} \{{r} \in R \wedge \mathit{explanandum} \in r.\mathit{actions}\}$

      \State  
      $\mathit{FiredRule} \gets$ \Find $r \in \mathit{CandidateRules~} \mathit{where} \hspace{1mm} \{$ \\
      $(\forall a \in r.\mathit{actions} : (a \in X \hspace{1mm} \wedge \funcall{time}(a) = \funcall{time}(explanandum))) \hspace{1mm} \wedge $\\
      $(\forall p \in r.\mathit{preconditions} : ({p} \in X \hspace{1mm} \wedge \funcall{time}(p) < \funcall{time}(explanandum)))\hspace{1mm} \wedge $ \\
        $  (\funcall{eval}(r.\mathit{preconditions}, X, \funcall{time}(explanandum)) = \textbf{true})  \} $

    \If {$\mathit{FiredRule} \neq \emptyset$} 
        \State $P  \gets \mathit{FiredRule} \wedge \mathit{FiredRule.preconds} \wedge \mathit{FiredRule.actions}$
    \EndIf
\State \Return \textit{$P$}

   \EndProcedure
   \end{algorithmic}
    }
    \label{algo:path}
\end{algorithm}

Subsequently, the algorithm searches through the rules to identify a subset of \textit{R}, which is referred to as \textit{CandidateRules}. This subset includes rules that contain the \explanandum as an action(\textit{{$r$}.actions} in Alg.~\ref{algo:path}, line 9). Next, the algorithm assesses each element of the \textit{CandidateRules} by traveling back through the log of events in $X$. The aim is to locate the specific rule \textit{r} that satisfies two conditions(lines 10-13): i) all of the other actions linked to \textit{r} must also have been triggered simultaneously with the \explanandum and ii), all of the preconditions associated with \textit{r} must have been satisfied before the \explanandum. To evaluate these conditions, it is sufficient to check the existence of the required preconditions and actions in $X$ and their timing order through a function that returns the timestamp of $X$. 

Please note that a rule may have multiple actions, which are all executed when the rule's required preconditions are met. The actions of rule \textit{r} are grouped using the \textit{AND} operator, meaning that \textbf{all} of \textit{r}'s actions must have been triggered. Unlike actions, preconditions of a rule can be combined using both the \textit{AND} and \textit{OR} operators. To determine the overall truth value of the rule's preconditions, the \textit{eval} function evaluates the logical expression formed by the logical operators of the preconditions. It takes preconditions of \textit{$r$}, along with $X$ and all timestamps, and outputs a Boolean value of \textit{True} if all necessary preconditions are met or \textit{False} otherwise. In short, it evaluates the innermost groups of preconditions first, assigning a truth value to them based on their operator. It then computes the truth value of the outer groups until the entire expression is resolved to a single truth value.

\subsubsection{\textbf{User-centric Explanation}}
\label{subsec:context}
Given Definition~\ref{def:psi}, a user-centric explanation necessitates the incorporation of \textit{AECs} coming from \AlExpComp, and \textit{CECs}. It generates a comprehensive explanation, which can be decomposed into multiple views to provide granular and personalized explanations (see Definition~\ref{def:Rep} %
Furthermore, the inclusion of a context-aware mechanism becomes imperative to determine the appropriate view for a given user in a specific situation (see Definition~\ref{def:inference} and Algorithm~\ref{algorithm:representation}).

In particular, we are introducing four different views for a user-centric explanation $\Psi$ in a rule-based system.
The first view %
is \textsf{Full Explanation View}, corresponds to the complete $\Psi$ that is composed of both \textit{AEC} (i.e., the path P) and \textit{CEC} (i.e., the rule definitions and the creators of rules). The subsequent views then offer partial representations of $\Psi$ by adjusting its granularity and length to cater to different user needs. In particular, the \textsf{Rule Explanation View} only denotes the rule $r_i \in R$, whose action needs explanation (i.e., the root of the path P). The \textsf{Fact Explanation View} displays the set of events that fulfill the preconditions of $r_i \in R$, resulting in the rule's activation. It provides a representation of the instances of the preconditions in the real world, thereby explaining why the rule has been activated. On the other hand, the \textsf{Simplified Explanation View} offers a high-level explanation for an \explanandum by simply stating that the event occurred due to a rule (without revealing the rule itself) set by a specific user $u_i$. 

\begin{figure}
  \centering
  \includegraphics[width=\linewidth]{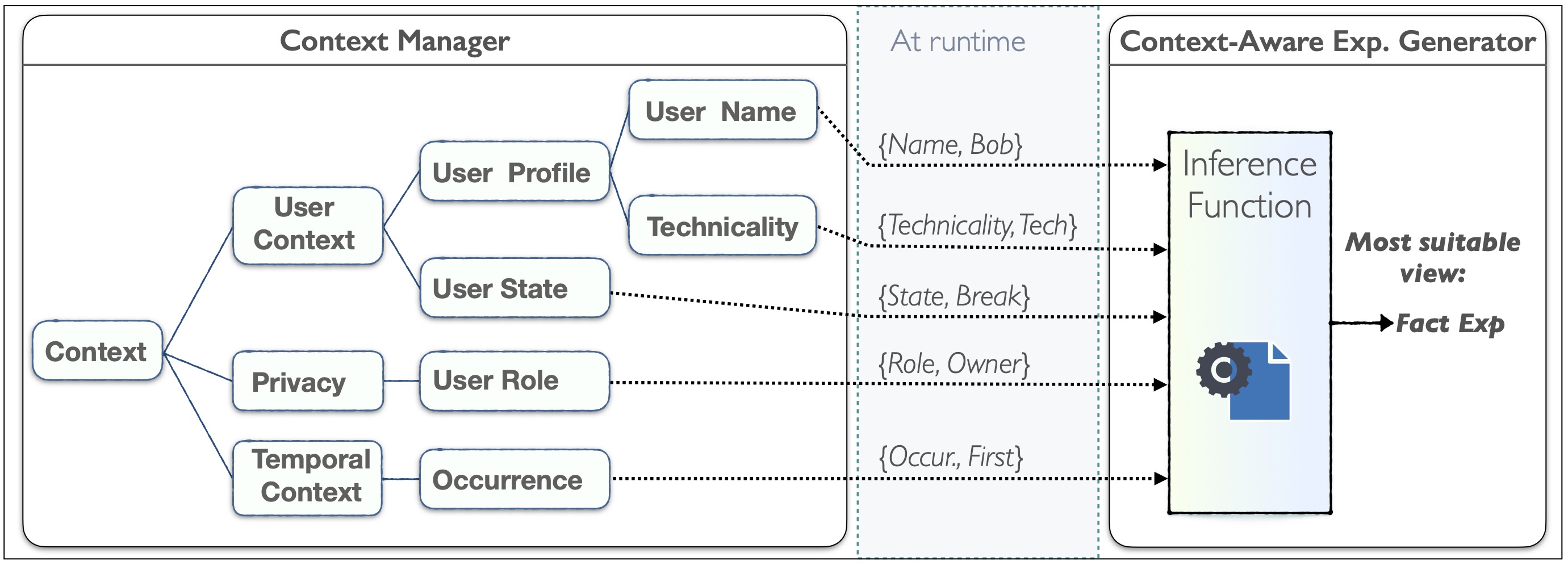}
  \caption{Context model (left) and its usage at runtime. The Context Manager provides the Inference Function with the current values of relevant contexts. In this example, the Inference Function infers that the most appropriate view to present to the user is the Fact Explanation View.}
  \label{fig:context}
\end{figure}

The final procedure to provide a context-aware and personalized explanation is to map a relevant set of contextual elements to the most suitable view. In our work, contexts are modeled as tuples of \textit{(attributeName, attributeValue)} pairs. We incorporate only relevant contexts for an explanation tailoring, meaning alterations in their value may impact the explanation's length and granularity. Thus, our context model is composed of a non-exhaustive set of contexts that we have found to be pertinent for preferring one explanation over another. More specifically, our system currently relies on some specification of three main general contexts (User, Privacy, and Temporal Context) as shown in Figure~\ref{fig:context}.

In our context model, \texttt{User Profile} captures static user attributes such as \texttt{User Name} and \texttt{Technicality Level}. The dynamic attributes of the user are modeled using the \texttt{User State}, which provides valuable insight into the user's current situation. For instance, if the \texttt{User State} is {\texttt{Working}}, it indicates that the user is busy and thus may require a shorter explanation. The frequency of exposure to a certain confusing situation, referred to as the \texttt{Occurrence} of the \explanandum in our model, is another significant factor affecting the level of detail required in an explanation. For example, if a user encounters a new \explanandum for the first time, they will naturally require a complete explanation compared to a situation where they have previously received an explanation for the same \explanandum. Finally, the \texttt{User Role} context accounts for the system's privacy settings, influencing the scope and type of information that can be disclosed in an explanation.

\begin{table}
\scriptsize
\centering
\caption{Context-view mappings.}
\label{tab:rep_policy}
\begin{tabular}{r|l|ccc|ccc|} 
\hline
\multirow{2}{*}{Expr} & \multirow{2}{*}{\diagbox{View }{Context}} & \multicolumn{3}{l|}{P1: User State}                                               & \multicolumn{3}{l|}{P2: Occurrence}                                                \\ 
\cline{3-8}
                      &                                                              & Mt\uline{}                & Bk                        & Wk\uline{}                & Ft                        & St                        & M                          \\ 
\hline\hline
1                     & Full Exp. View                                               & x                         & \checkmark & x                         & \checkmark & x                         & x                          \\
2                     & Fact Exp. View                                               & x                         & \checkmark & \checkmark & \checkmark & \checkmark & x                          \\
3                     & Rule Exp. View                                               & \checkmark & \checkmark & \checkmark & x                         & \checkmark & x                          \\
4                     & Simplified Exp. View                                         & \checkmark & \checkmark & \checkmark & x                         & x                         & \checkmark  \\
\bottomrule
\bottomrule
  &                       & \multicolumn{3}{l|}{P3: Technicality}                & \multicolumn{3}{l|}{P4: Role}   \\ \cline{3-8}
  &                       & Tch         & Med        & Ntch                & Ow          & Cw         & Gst                         \\ \hline\hline
1 & Full Exp. View        & \checkmark  & \checkmark & x                   & x           & \checkmark & x                          \\
2 & Fact Exp. View        & \checkmark  & x          & x                   & \checkmark  & \checkmark & x                          \\
3 & Rule Exp. View        & \checkmark  & \checkmark & x                   & \checkmark  & \checkmark & x                          \\
4 & Simplified Exp. View  & \checkmark  & x          & \checkmark                   & \checkmark  & \checkmark & \checkmark  \\

\bottomrule

\multicolumn{8}{l}{  \begin{tabular}[c]{@{}l@{}} \scriptsize \textbf{Expr:} {Expressiveness}, \textbf{Mt:} {Meeting},  \hspace{1pt} \textbf{Bk:} { Break},  \hspace{1pt}\textbf{Wk:} { Working},   \hspace{1pt}\textbf{Ft:} { First Time}, \\  \scriptsize \textbf{St:} { Second Time}, \hspace{1pt}\textbf{M:} {More},  \textbf{Tch:} {Technical}, \textbf{Med:} {Medium Technical}, \hspace{1pt} \textbf{Ntch:}\\ \scriptsize {Not Technical},  \hspace{1pt}\textbf{Ow:} {Owner},\hspace{1pt}\textbf{Cw:} {Co-Worker}  \hspace{1pt}\textbf{Gst:} {Guest}. The \checkmark symbols: the view \\  \scriptsize fits for the context variable. An x symbol: the view is not suitable that context. \end{tabular}}

\end{tabular}
\end{table}

The \ConMangComp is responsible for monitoring and retrieving current contextual element values from various sources.%
It supplies the \ConExpComp with two types of data. Firstly, the relevant contextual information to build $\Psi$ (i.e., \textit{CEC}) and secondly, the relevant contexts to infer the best-suited explanation view for the given situation. As illustrated in Figure~\ref{fig:context}, the current values of contextual elements are provided to the \textit{Inference Function} by the \ConMangComp during the explanation generation process. The \textit{Inference Function} uses this information to infer the appropriate explanation view to generate. This is done via Algorithm \ref{algorithm:representation}, utilizing contextual elements as input to determine the most suitable \textit{View} for the situation.

Specifically, we have implemented a set of mapping policies using rules in a rule engine. Table~\ref{tab:rep_policy} shows a simplified representation of them. The priorities (P1 to P4) for each context determine the evaluation order. The table shows what the suitable views are for each possible value of a given context (denoted by $\checkmark$ symbol). The table also represents the expressiveness order of views where the \textit{Full Explanation} is the most expressive one followed by, \textit{Fact}, \textit{Rule}, and \textit{Simplified}.

Algorithm~\ref{algorithm:representation} shows the process by which the \textit{Inference Function} selects the appropriate view. Using the Context-View mappings and the expressiveness order of views, it calculates a \textit{$view$} from the set of \textit{Views} that is suitable for the given situation. At the outset of the process, the \textit{Inference Function} initializes the set of available \textit{Views} to include all possible types of views (line 4). It then sorts the policy rules in order of priority, starting with the highest (line 7). The function then begins executing each policy by evaluating the conditions specified in the rules, that is, by assessing the actual values of the relevant contexts at that time. The result of each policy execution is a set of so-called \textit{Suitable$\underline{}$View} (line 9). For instance, suppose a user, denoted as \textit{U1}, is on a break, then the corresponding suitable views are \textit{[Full, Fact, Rule, and Simplified]}. Then, the algorithm computes the intersection between the set of all views and the set of \textit{Suitable$\underline{}$View} for \textit{U1}. If the intersection is non-empty, the algorithm replaces the \textit{Views} set with the resulting set (lines 10--11).

\begin{algorithm}
\scriptsize
    \caption{Inference Function}
    
    \begin{algorithmic}[1]
    \Procedure{\funcall{find most suitable view for given contexts}}{}\\ 
   \textbf{input:} {Set of Context\underline{ }View mappings, Expressiveness order of View, Current values of contextual elements }  \\
   \textbf{output:} {suitable $\mathit{view} \in \mathit{Views}$}

    \State $\mathit{Views} \gets \{\mathit{Full Exp}, \mathit{Fact Exp}, \mathit{Rule Exp}, \mathit{Simplified Exp}\}$
    \State $\mathit{Policy}\gets$ Set of all the $\mathit{Context\_View}$ mappings \algorithmiccomment{\textcolor{gray}{see Table~\ref{tab:rep_policy}}}
    
    \State {$\mathit{Context\_Knowledge}\gets \{({C_1},\mathit{Value}_1) , \ldots ,  ({C_i},\mathit{Value}_i)$\}}
    
     \State {\textbf{sort}  $P \in \mathit{Policy}$ from highest priority to lowest} 
    
    \For {${P} \in \mathit{Policy}$} 
    \State $\mathit{Suitable\_View} \gets \mathit{\funcall{apply}}(\mathit{Policy}, \mathit{Context\_Knowledge})$

          \If {$\mathit{Views} \cap \mathit{Suitable\_View} \neq\varnothing$} 

        \State $\mathit{Views}\gets \mathit{Views} \cap \mathit{Suitable\_View}$
    \EndIf

       \EndFor

\State {\textbf{sort}  $\mathit{view} \in \mathit{Views}$ based on expressiveness} 

\State \Return most expressive $\mathit{view} \in \mathit{Views}$

   \EndProcedure
   \end{algorithmic}
    
    \label{algorithm:representation}
\end{algorithm}

Afterward, the algorithm repeats the same process for policies with the next priority, which in this case is the \textit{Occurrence} context. Assume that \textit{U1} is requesting an explanation for a particular \explanandum for the \textit{second} time. Based on the policy for the \textit{Occurrence} context in Table~\ref{tab:rep_policy}, the set of suitable views will exclude the \textit{Full} and \textit{Simplified Explanation}. The decision-making heuristic applied here is based on the fact that the \textit{Full Explanation} view is deemed too detailed and is best suited for first-time explanations, while the \textit{Simplified Explanation} view is too abstract and is better suited for situations where a user has faced the \explanandum more than twice in the last three months. 
Consequently, the set of \textit{Views} at this stage is restricted to \textit{Rule} and \textit{Fact Explanation}. The procedure then continues by examining the rules for the remaining contexts. As shown in Table~\ref{tab:rep_policy}, our implemented \exen adheres to the context model presented in Section~\ref{subsec:rul} and includes \textit{User Technicality} and \textit{User Role} in addition to \textit{User State} and \textit{Occurrence} discussed earlier. In brief, for the \textit{Technicality} and \textit{User Role} contexts, the Context-View mappings assign the simplest and most abstract explanation's view to \textit{Non-technical} and \textit{Guest} users.

The last step is the sorting of the \textit{Views} set based on their expressiveness by the \textsl{Inference Function}. The most expressive view is then returned as the final output. Note that the check at line 10 ensures that \textit{Views} set will never be empty. The \textsl{Explanation Presentation} Component is responsible for translating this output to natural language through the \textsl{Transformation Function}, which could be implemented via different approaches, such as a template-based text generation or Prompt-Based Text Generation via Large Language Models. 
 
\section{Implementation and Feasibility Check}\label{Sec:imp}
\begin{figure*}
  \centering
  \includegraphics[width=0.7\linewidth]{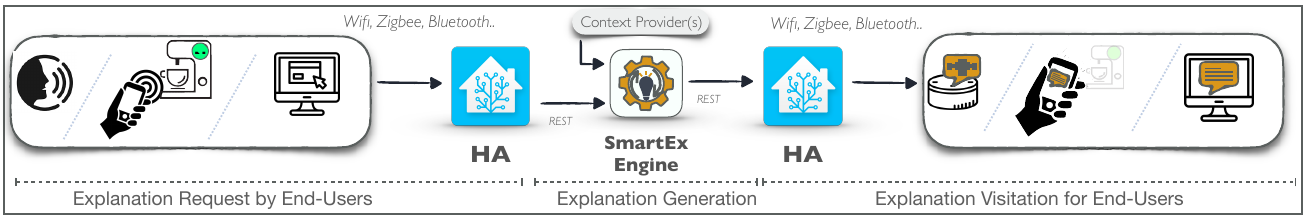}
  \caption{Explanation generation workflow}
  \label{fig:exWorkflow}
  \vspace{-1em}
\end{figure*}

\begin{table*}
    \centering
    \caption{Explanations of the same \explanandum (TV is muted) for three different users (Bob, Alice, Dana)}
    \label{tab:s1}
    \begin{tabularx}{\textwidth}{@{}lp{4cm}lX@{}}
    \toprule
    \textbf{User} & \textbf{Context (CEC)} & \textbf{View} & \textbf{Explanation}\\
    \midrule
    Bob & \{State: Break, Occurrence: 1st time, Technicality: Tch, Role: \textbf{Ow}\}  & 
    Fact Expl. & Hi Bob, tv\_mute is active because currently a meeting in room 1 is going on and the TV is playing.\\
    Alice &  \{State: Break, Occurrence: 1st time, Technicality: Tch, Role: \textbf{Cw}\} & Full Expl. & Hi Alice, tv\_mute is active because Bob has set up a rule: ``Rule\_2: mutes the TV if the TV is playing while a meeting is going on'' and currently a meeting in room 1 is going on and the TV is playing, so the rule has been fired.\\
    Dana & \{State: Break, Occurrence: 1st time, Technicality: Tch, Role: \textbf{Gst}\}   & Simpl.\ Expl. & Hi Dana, Bob has set up a rule and at this moment, the rule has been fired.\\
    \bottomrule
    \vspace{-1.5em}
    \end{tabularx}
\end{table*}
We created a prototype implementation as a proof of concept and tested it extensively in various scenarios in our own \textit{\smartEnv lab}. Due to space constraints, we report results for only one scenario in this paper. 
The prototype is a valuable basis for conducting user studies that we plan for future research.

We have developed \exen \footnote{\url{https://github.com/ExmartLab/SmartEx-Engine}} as a RESTful web service implemented in Java, using MongoDB as a database. This web service can be integrated into existing non-explainable systems to create an explainability layer on top of them. In our lab, we have an array of sensors, actuators, smart lights, plugs, and appliances working on WiFi, Zigbee, and Bluetooth. Home Assistant (HA)\footnote{\url{https://www.home-assistant.io/}} serves as our software hub, enabling task automation via its rule-based mechanism.
HA offers a RESTful API with endpoints for fetching data, including user information, automation rules, device states, and a log of past activities and states. The \ConMangComp in \exen fetches this runtime data continuously from HA, while more static data, such as the user profiles, is directly stored in the \exen database. Collecting more contextual information about the user's state (e.g., meeting) must be provided by additional services, which are not part of \exen. Additionally, the \ConMangComp stores situations that have already been explained to a user in the last three months. This information is then used by the \ConExpComp to tailor the explanation accordingly (based on the \texttt{Occurrence} context).
The \textsl{Inference Function}, which determines a suitable view of the explanation based on the current context, has been implemented using the Easy-Rule\footnote{\url{https://github.com/j-easy/easy-rules}} 
Java rule engine. Accordingly, the View-Context mappings policies (see Table~\ref{tab:rep_policy}) are defined as rules processable by the rule engine. 
Lastly, the determined view is sent to the \textsl{Explanation Presentation} component, which utilizes a \textit{Transformation Function} to generate natural language, using a template-based approach for English text generation.

The workflow in Figure~\ref{fig:exWorkflow} displays how users can seek explanations for unexpected behavior using voice commands (e.g., 'What just happened?'), NFC tag scans, or a custom HA dashboard. The first option offers a general explanation of the latest system action, while the latter enables users to specify a particular device's action. Explanations are delivered via audio and visual media, read aloud by the voice controller, and displayed on client applications such as mobile apps (Android/iOS) and web browsers

\textbf{Exemplary Scenario: TV suddenly mutes.} To demonstrate our prototype, we pick the motivating scenario described in Section~\ref{subsec:Motiv} where the kitchen TV is muted during meetings in nearby rooms to prevent disruption, involving three users: Alice, Bob, and Dana.  

To generate explanations, \exen collects \textit{AEC} (see Section~\ref{subsec:algorithmic}) and \textit{CEC} (see Section~\ref{subsec:context}). The AECs are context and user-independent. The algorithm identifies that the reason for muting the TV is that a specific rule, called \textit{Rule\_2}, was fired. \exen then uses the collected CECs to generate a context-aware explanation. 
Table~\ref{tab:s1} shows the explanations generated by \exen for the same \explanandum (i.e., TV is muted) for the three individual users displayed on their devices.

To generate a context-aware and personalized explanation for the three users, the \ConExpComp must determine the best-suited \textit{View} based on the CECs and the rules specified in Table~\ref{tab:rep_policy}. The \ConExpComp determines the \textit{Fact Exp.} for Bob, the \textit{Full Exp.} for Alice, and the \textit{Simplified Exp.} for Dana.

\section{Conclusion}\label{Sec:con}
Despite theoretical foundations and empirical findings, there's a notable lack of effective software engineering methods and technical solutions for generating user-centric explanations in smart environments. To bridge this gap, we proposed a novel mechanism and reference architecture for explanation generation in this domain. The framework generates context-aware and personalized explanations on various levels of granularity tailored to users and situations. The framework follows a service-oriented architecture that enables integration with existing pervasive smart environment systems to equip them with self-explainability functions. Our experiments on generating explanations for everyday scenarios in smart environments show the feasibility of the approach and motivate further work.  In forthcoming research, we plan to expand and enrich our context model to encompass a broader range of contextual elements. Furthermore, we plan to incorporate other types of intelligent systems (recommendation systems, ML-based systems, etc.) and more complex types of explanations, such as contrastive explanations~\cite{Herbold24}. Also, we plan to make the explanation process more interactive, so users can provide feedback to the system to receive more detailed explanations.

\bibliographystyle{IEEEtran}
\bibliography{ref}

\end{document}